\documentstyle[12pt]{article}
\begin{document}
%
%
\begin{center}
{\bf Non-Resonant Effects in Implementation of Quantum Shor Algorithm}\\ \ \\
G.P. Berman$^1$, G.D. Doolen$^1$, G.V. L\'opez$^2$, and V.I. Tsifrinovich$^3$\\ \ \\
\end{center}
$^1$Theoretical Division and CNLS, \\
Los Alamos National Laboratory, Los Alamos, New Mexico 87545\\
$^2$ Departamento de F\'isica, Universidad de Guadalajara,
Corregidora 500, S.R. 44420, Guadalajara, Jalisco, M\'exico\\
$^3$Department of Physics, Polytechnic University,\\
Six Metrotech Center, Brooklyn NY 11201\\ \ \\
\begin{center}
{\bf ABSTRACT}
\end{center}
We simulate Shor's algorithm on an Ising spin quantum computer. The influence 
of non-resonant effects is analyzed in detail. It is shown that 
our ``$2\pi k$''-method successfully suppresses non-resonant effects 
even for relatively large values of the Rabi frequency.
\newpage
\quad\\
{\bf I. The Quantum Shor Algorithm}\\ \ \\
The quantum Shor algorithm \cite{1} provides an exciting opportunity for prime-factorization of large integers -- a problem beyond the capabilities of today's powerful digital computers. Shor's algorithm utilizes two quantum registers (the $x$- and $y$- registers), which contain quantum bits 
two-level quantum systems called qubits \cite{1}-\cite{3}. First, the  quantum computer produces the uniform superposition of all states in the $x$-register -- all possible values of $x$. Second, the quantum computer computes the periodic function: $y(x)=q^x (mod~N)$, where $N$ is the number to factorize, and  $q$ is any number which is coprime to $N$. Third, the quantum computer creates a discrete Fourier transform of the $x$-register. The measurement of the state of the $x$-register yields the 
period, $T$, of the function $y(x)$, which is used to produce a factor of the number $N$. 

In Dirac notation, the wave function of the quantum computer can be represented as a superposition of digital states,
$$
|a_{L-1}a_{L-2}...a_1a_0,b_{M-1}b_{M-2}...b_1b_0\rangle,\eqno(1)
$$
where $a_k$ ($0\le k\le L-1$) denotes the state of the $k$-th qubit in the $x$-register, and  $b_n$ ($0\le n\le M-1$) denotes the state of the $n$-th qubit in the $y$-register. For example, if the $k$-th qubit of the $x$-register is in the ground state, then: $a_k=0$, and if it is in the excited state, $a_k=1$.

In decimal notation, the digital state can be represented as $|x,y\rangle$, where
$$
x=\sum_{k=0}^{L-1}a_k2^k,~y=\sum_{n=0}^{M-1}b_n2^n.\eqno(2)
$$
In this notation, the initial wave function of a quantum system is: $|0,0\rangle$. The uniform superposition of the states created in the $x$-register can be written as,
$$
\Psi={{1}\over{\sqrt{D}}}\sum_x|x,0\rangle,\eqno(3)
$$
where $D=2^L$ is the number of states in the $x$-register. After computation of the function $y(x)$, we have,
$$
\Psi={{1}\over{\sqrt{D}}}\sum_x|x,y(x)\rangle.\eqno(4)
$$
After the discrete Fourier transform, one measures the state of the $x$-register. The probability of the measurement, $P(x)$, must be a peaked distribution with peak separation, $\Delta x$, equal to a multiple of $1/T$. In particular, if the number of states, $D$, in the $x$-register is divisible by the period, $T$, then: $\Delta x=D/T$. From the value $\Delta x$ one can find the period, $T$. A factor of the number $N$ can be found by computing the greatest common divisor of $(q^{T/2}+1)$ and $N$, or  $(q^{T/2}-1)$ and $N$
(for even $T$). 

It was shown in \cite{4}, that the simplest demonstration of the quantum Shor's algorithm can be done with only four qubits. (Two qubits represent the $x$-register and two qubits represent the $y$-register.) This primitive quantum computer is able to find a factor of the number $4$. For $N=4$, the only coprime number is $q=3$. The function, $y(x)=3^x (mod~4)$, has only two values: $y=1$ and $y=3$, and a period is, $T=2$. In Dirac notation, the wave function (4) has the form,
$$
\Psi={{1}\over{2}}(|00,01\rangle+|01,11\rangle+|10,01\rangle+|11,11\rangle).
\eqno(5)
$$
After the Discrete Fourier transform, the wave function of the quantum computer is,
$$
\Psi={{1}\over{2}}(|00,01\rangle+|00,11\rangle+|10,01\rangle+|10,11\rangle).
\eqno(6)
$$
Measuring $x$ can produce two values: $x=0$ and $x=2$. So, $\Delta x=2$, and the period is: $P=D/\Delta x=2$. Finally, $q^{P/2}-1=3-1=2$, and the factor of $4$ can be found as the greatest common divisor of $4$ and $2$.\\ \ \\
{\bf 2. The Ising Spin Quantum Computer}\\ \ \\
The Ising spin quantum computer, first introduced in \cite{5}, consists of a one-dimensional chain of $1/2$ spins, connected by Ising interactions. The quantum logic gates can be implemented in the Ising spin quantum computer using selective electro-magnetic pulses which induce transitions between the ground state and the excited state of a chosen spin. The Hamiltonian of the quantum computer during the action of the $n$-th electro-magnetic pulse can be written as,
$$
{\cal H}=-\sum^{S-1}_{k=0}\Bigg\{\omega_kI^z_k+{{1}\over{2}}\Omega_{kn}
\Bigg(e^{-i(\omega^nt+\varphi_n)}I^-_k+e^{i(\omega^nt+\varphi_n)}I^+_k\Bigg)\Bigg\}-\eqno(7)
$$
$$
2\sum_{k=0}^{S-2}J_{k,k+1}I^z_kI^z_{k+1},
$$
where $\omega_k$ is the resonant frequency of the $k$-th spin; $\Omega_{kn}$ is the Rabi frequency of the $k$-th spin during the $n$-th pulse; $\omega^n$ and $\varphi_n$ are the frequency and the phase of the $n$-th electro-magnetic pulse; $J_{k,k+1}$ is the constant of the Ising interaction, $I^z_k$ and $I^{\pm}_k=I^x_k\pm iI^y_k$ are the operators for the $k$-th $1/2$ spin; $S$ is the total number of spins (qubits) in the quantum computer, and we choose units in which $\hbar=1$.  
 The difference of resonant frequencies between spins (qubits) can be provided by differences in the gyromagnetic ratio of the spins or by a non-uniform external magnetic field. The Ising interactions also influence the resonant frequency, providing an opportunity for conditional logic gates. The main disadvantage of the Ising spin quantum computer is the presence of non-resonant effects: a selective electro-magnetic pulse which is resonant to a specific spin, influences all other spins of the quantum computer.

Destructive non-resonant effects can be significantly weakened by using a $2\pi k$-method suggested in \cite{6} (see also \cite{3}, Chapter 22). The main idea of this method is that parameters of the electro-magnetic pulse should be chosen in such a way that a non-resonant spin rotates about the effective magnetic field by the angle of $2\pi k$ (where $k$ is an integer). So, at the end of the pulse, it returns to its initial position.

To the best of our knowledge, this paper provides the first simulation of the quantum Shor's algorithm, taking into consideration non-resonant effects. We investigate the destructive influence of non-resonant effects and show that one can significantly reduce non-resonant effects by using $2\pi k$-pulses.\\ \ \\
{\bf 3. Simulation of Quantum Computation}\\ \ \\
We describe the 4-qubit quantum computer with the Hamiltonian (7) putting $S=4$, $\omega_{k+1}=\omega_k+\Delta\omega$, $\Omega_{kn}=\Omega$, $J_{k,k+1}=J$. We count spins (qubits) from the right to the left. The values of parameters will be given below. To implement the Shor's algorithm we consider application of 16 selected resonant pulses. The Hamiltonian (7) suggests that the $n$-th pulse is ``cut'' from a continuous harmonic oscillation, $\exp(i\omega^nt)$ and its phase is shifted by $\varphi_n$.
For example, the first pulse has frequency, $\omega^1=\omega_2+2J$, and  phase, $\varphi_1=\pi/2$; the second pulse has frequency, $\omega^2=\omega_3+J$, and phase, $\varphi_2=\pi/2$, and so on.  We will use single-integer decimal notation for the $4$-spin basic states, {\it i.e.},
$$
|a_1a_0,b_1b_0\rangle=|p\rangle,~p=b_0+2b_1+2^2a_0+2^3a_1,
$$
and $c_p$ is the amplitude corresponding to the state $|p\rangle$. The Schr\"odinger equation for the amplitude, $c_p$, of the state, $|p\rangle$, can be written as,
$$
 i\dot c_p=\sum_{m=0, m\neq p}^{15}c_mV_{pm}\exp{[i(E_p-E_m)t+r_{pm}(\omega^nt+\varphi_n)]}.\eqno(8)
$$
Here we use the interaction representation, {\it i.e.} $c_p\rightarrow c_p\exp(-iE_pt)$; $V_{pm}$ is the matrix element between the states $|p\rangle$ and  $|m\rangle$ which is equal to $V_{pm}=-\Omega/2$ for single-spin transitions and zero for other transitions; $E_p$ and $E_m$ are the energies of the corresponding states, $r_{pm}=-1$ for $E_p>E_m$, and  $r_{pm}=1$ for $E_p<E_m$.  If one neglects non-resonant effects, the $n$-th pulse produces only a rotation of a resonant spin which can be described as a transformation of the basic states:
$$
|0\rangle\rightarrow \cos(\alpha_n/2)|0\rangle+i\exp(-i\varphi_n)\sin(\alpha_n/2)|1\rangle,\eqno(9)
$$
$$
|1\rangle\rightarrow \cos(\alpha_n/2)|1\rangle+i\exp(i\varphi_n)\sin(\alpha_n/2)|0\rangle,
$$
 where $\alpha_n=\Omega\tau_n$ is the angle of rotation of the average spin, and $\tau_n$ is the duration of the $n$-th pulse. For example, first two pulses are the $\pi/2$-pulses, {\it i.e.} $\alpha_1=\alpha_2=\pi/2$. 

Instead of a direct discrete Fourier transform we have used an idea of Coppersmith and Deutsch, first described in \cite{7}. Following this idea, the final wave function (without non-resonant effects) contains four states:
$$
|00,01\rangle,~|00,11\rangle,~|01,01\rangle,~|01,11\rangle.\eqno(10)
$$
One must reverse the results of the measurement of the $x$-register to get the actual result of the discrete Fourier transform. Following this rule, one obtains: $x=0$ and $x=2$ with equal probability, $1/2$.

It follows from (10) that in the resonant approximation we have,
$$
|c_1|^2=|c_3|^2=|c_5|^2=|c_7|^2=1/4,\eqno(11)
$$
and zero probabilities for all other states.
\\ \ \\
{\bf 4. Results of Simulations of the Shor's Algorithm}\\ \ \\

Fig. 1a, shows the actual values of the ``resonant'' probabilities (11) for the following values of parameters,
$$
\Delta\omega=10,~J=1,~\Omega=0.1.\eqno(12)
$$
One can see that the expression (11) is approximately satisfied. Small deviations of the resonant probabilities from 1/4 are associated with nonzero probabilities for ``non-resonant'' states. Fig. 1b, 
(with a magnified scale) shows that non-resonant states are very non-uniformly excited: the probability of the upper state, $|15\rangle$, is close to $10^{-3}$; the probabilities of the states $|9\rangle$ and $|11\rangle$ are close to $10^{-4}$; and all other non-resonant probabilities have smaller values.
Fig. 2 shows a radical change of the results for a small increase of the Rabi frequency (from $\Omega=0.1$ to $\Omega=0.112$). Now the probabilities of the resonant states are significantly different: the probability of measuring 
$x=2$ is noticeably greater than for the value $x=0$. The probabilities of error connected with non-resonant states has sharply increased. 

Fig. 3 shows the return to the quasi-resonant picture for $\Omega=0.125$: the probabilities of the resonant states (9) are close to 1/4, and the probabilities of the non-resonant states are less than $10^{-3}$. Figs 4 through 7 demonstrate the same periodicity at increasing the Rabi frequency: ``quasi-ideal'' implementation of Shor's algorithm at $\Omega=0.1666$, $\Omega=0.25$ (Figs 5 and), and its destruction for intermediate values, $\Omega=0.1458$, $\Omega=0.2083$ (Figs 4 and 6). Fig. 8 demonstrates the dependence of resonant probabilities on the value of the Rabi frequency, $\Omega$, which continuously approaches the value of the Ising constant, $J$. The resonant probabilities are close to (11) at fixed values of $\Omega$:
$$
\Omega=0.1,~0.125,~0.1666,~0.25.\eqno(13)
$$
The last point where the probabilities approach the value 1/4 is: $\Omega=0.51639$, but the deviation from 1/4 at this point is much greater than for previous values of $\Omega$ (13). Fig. 9 demonstrates the destruction of the Shor's algorithm when the value of the frequency difference between neighboring qubits, $\Delta\omega$, approaches the value of the Ising constant, $J$. (The Rabi frequency, $\Omega$ and $J$ are given by (12).) Significant deviation of
the probabilities from their ``resonant'' values begins at approximately  $\Delta\omega=3$. The greatest probability of error occurs in the upper state, $|15\rangle$. Fig. 10 shows the time dependence of the resonant probability, $|c_3|^2$, during the action of the last two pulses, for different values of $\Delta\omega$. One can see, that for $\Delta\omega\ge 4$, the function $|c_3(t)|^2$ roughly does not depend on $\Delta\omega$. For $\Delta\omega\le 3$, this function depends significantly on $\Delta\omega$.

Note that, in Figs 1-10, the angles of rotation, $\alpha_n$, did not change, {\it i.e.} the increase of $\Omega$ was compensated by the decrease of the pulse dutation, $\tau_n$. Fig. 11 shows the destruction of Shor's algorithm when the phases, $\varphi_n$, deviate randomly from their correct values, $\varphi_n^0$. The phases were chosen randomly in the interval $(\varphi_n^0-\varepsilon,\varphi_n^0+\varepsilon)$, for four values of $\varepsilon$. Significant influence of the phase fluctuations can be observed at $\varepsilon\ge 0.5$ rad. Again, the main error occurs in the upper state, $|15\rangle$. Fig. 12 shows the influence  of the fluctuation of the angle of rotation, $\alpha_n$ on the performance of Shor's algorithm. We have randomly changed the values of $\tau_n$ in the interval $(\tau_n^0-\varepsilon,\tau_n^0+\varepsilon)$, for different values of $\varepsilon$. The values of $\Omega$, $\Delta\omega$, and $\tau_n^0$ are given by (12), where $\tau_n^0$ is the correct value of $\tau_n$. The noticeable destruction in the performance of Shor's algorithm was observed at $\varepsilon\ge 2$. The corresponding angle of rotation is: $\Omega\varepsilon\ge 0.2$ rad.\\ \ \\
{\bf 5. Discussion}\\ \ \\
The most important result of our simulation of Shor's algorithm is the successful periodic suppression of non-resonant effects when the value of Rabi frequency, $\Omega$, approaches the Ising constant, $J$. We shall now discuss this phenomenon in detail.

Suppose the $n$-th electro-magnetic pulse induces a non-resonant transition between the states $|p\rangle$ and $|m\rangle$, where $E_p>E_m$. The equation of motion (8) for these two states can be written approximately as,
$$
i\dot c_p=-(\Omega/2)c_m\exp\{i[(E_p-E_m-\omega^n)t-\varphi_n]\},\eqno(14)
$$
$$
i\dot c_m=-(\Omega/2)c_p\exp\{i[(E_m-E_p+\omega^n)t+\varphi_n]\}.
$$
Here we assume that the probability of the non-resonant transition between the states $|p\rangle$ and $|m\rangle$  is much greater than any other transition probability involving these two states. With an accuracy up to the phase, the solution of (14) can be written as,
$$
c_p(t_n)=c_p(t_{n-1})[\cos(\Omega_e\tau_n/2)-(i\Delta/\Omega_e)\sin(\Omega_e\tau_n/2)]+c_m(t_{n-1})[(i\Omega/\Omega_e)\sin(\Omega_e\tau_n/2)],\eqno(14a)
$$
$$
c_m(t_n)=c_m(t_{n-1})[\cos(\Omega_e\tau_n/2)+(i\Delta/\Omega_e)\sin(\Omega_e\tau_n/2)]+c_p(t_{n-1})[(i\Omega/\Omega_e)\sin(\Omega_e\tau_n/2)].
$$
Here $\Omega_e$ is the effective frequency for the non-resonant transition,
$$
\Omega_e=(\Omega^2+\Delta^2)^{1/2},\eqno(15)
$$
and $\Delta=E_p-E_m-\omega^n$, $(t_{n-1},t_n)$ is the time-interval of the $n$-th pulse; $t_n-t_{n-1}=\tau_n$.

One can see that for $\Omega_e\tau_n=2\pi k$ $(k=1,2,...)$ the probability of non-resonant transition vanishes. This is the main idea of the ``$2\pi k$'' method \cite{3,6}. For a $\pi$-pulse ($\Omega\tau_n=\pi$), the values of $\Omega$ which satisfy this $2\pi k$-condition are,
$$
\Omega=|\Delta|/\sqrt{4k^2-1}=|\Delta|/\sqrt{3},~|\Delta|/\sqrt{15},~|\Delta|/\sqrt{35},~
|\Delta|/\sqrt{63},...\eqno(16)
$$
For a $\pi/2$-pulse, the corresponding values of $\Omega$ are,
$$
\Omega=|\Delta|/\sqrt{16k^2-1}=|\Delta|/\sqrt{15},~|\Delta|/\sqrt{63},...\eqno(17)
$$
If the Rabi frequency satisfies the $2\pi k$-condition (17) for a $\pi/2$-pulse, it automatically satisfies the condition (16) for a $\pi$-pulse.

Let us consider the most probable non-resonant transitions in our system. We assume that $J\ll\Delta\omega$, so the most probable non-resonant transitions occur when the frequency difference, $|\Delta|$, is of the order of $J$. For the right spin ($k=0$) and the left spin ($k=3$) we have two frequencies depending on the state of the only neighbor: $\omega_k+J$ and $\omega_k-J$. 
 If the {\it rf} pulse is turned on either of these transitions, the frequency difference, $|\Delta|$, for the other one will be $2J$. For the inner spins $(k=1,2)$, we have three possible resonant frequencies: 
$\omega_k$, $\omega_k-2J$ and $\omega_k+2J$. The frequency difference, $|\Delta|$, can take two values: $2J$ and $4J$. 

One cannot suppress both non-resonant transitions with $|\Delta|=2J$ and $|\Delta|=4J$. Putting $|\Delta|=2J$ in (17) we obtain,
$$
\Omega=2J/\sqrt{16k^2-1}\approx J/2k.
$$
For $|\Delta|=4J$, we have, 
$$
\Omega=4J/\sqrt{16k^2-1}\approx J/k.
$$
So, if one satisfies the $2\pi k$-condition for $|\Delta|=2J$, one satisfies approximately the same condition for $|\Delta|=4J$. But total suppression of two non-resonant transitions is impossible. 

Now we consider an example -- the suppression of non-resonant effects during the first two pulses. The initial state of our system is the ground state, $|00,00\rangle$. First, we apply a $\pi/2$ pulse with the frequency, $\omega_2+2J$, which is resonant with the spin with the number $k=2$. The non-resonant effects for three other spins are small if $\Delta\omega$ is large enough. After the action of the first pulse, we have the superposition,
$$
(1/\sqrt{2})(|00,00\rangle+|01,00\rangle).\eqno(18)
$$
Next, we apply a $\pi/2$-pulse with the frequency $\omega_3+J$. This pulse is resonant with the  $k=3$ spin (in its ground state, the first term in (18)).
 But this pulse also perturbs the same spin in the excited state (the second term in (18)). The frequency of the transition for the spin with $k=3$ in the second term in (18) is $\omega_3-J$. So, the frequency difference is: $|\Delta|=2J$. Taking a value of $\Omega$ which satisfies  (17) with $|\Delta|=2J$, one suppresses this non-resonant transition.

Putting $\Omega=2J/\sqrt{16k^2-1}$ with $J=1$ we obtain the values:
$$
\Omega\approx 0.100, 0.125, 0.167, 0.252, 0.516, 
$$
which are close to the values which were obtained from the computer simulations.

To decrease the time of quantum computation, one should use the largest possible values of $\Omega$. If the acceptable probability of error state is of the order of $10^{-3}$, then one should use the value $\Omega\approx 0.25$ (Fig. 7). If a probability of the error state of $10^{-2}$ is acceptable, one can use the value, $\Omega\approx 0.516$. The time of quantum computation can be further reduced if one uses different values of $\Omega$ for $\pi/2$- and $\pi$- pulses. According to (16) and (17), the maximum value of the Rabi frequency, $\Omega$, for a $2\pi$ rotation of non-resonant spins is,
$\Omega\approx 0.51639$ for $\pi/2$-pulses, and $\Omega\approx 1.1547$ for $\pi$-pulses. Fig. 13 shows the probabilities of the states for this minimal-time Shor's algorithm implementation. The greatest probability of error $\sim 10^{-2}$ is connected with the states $|11\rangle$ and 
$|15\rangle$. 
Note, that for $\pi$-pulses the value of Rabi frequency is greater than $J$ and the $2\pi k$-method still suppresses the non-resonant effects.\\ \ \\
{\bf Conclusion}\\ \ \\
We reported the first simulation of the quantum Shor's algorithm for prime factorization with taking
into consideration non-resonant effects. We have considered the Ising spin quantum computer of four qubits: two qubits in the $x$-register, and two qubits in the $y$-register. This primitive quantum computer is able to factor the smallest composite number, $N=4$. While a simulation of much more complicated system of 15 qubits was reported in {\cite{9}, that work deals with well-separated ions in an ion trap and it does not take into consideration non-resonant effects.

We have studied the destructive influence of non-resonant effects when the Rabi frequency, $\Omega$, approaches the value of the Ising interaction constant, $J$. We also studied the influence of the random fluctuations of the duration and the phase of the electro-magnetic pulses. 

The main results of our consideration are the following:

1. When the Rabi frequency, $\Omega$, approaches the Ising interaction constant, $J$, there is no monotonic increase of destructive non-resonant effects. Non-resonant effects are suppressed effectively when the value of $\Omega$ satisfies,
$$
\Omega=2J/\sqrt{16k^2-1},~k=1,2,....\eqno(19)
$$
for all pulses.
These values correspond to $2\pi k$ rotation of a non-resonant spin whose frequency is the most close to the resonant frequency. This method of suppression of non-resonant effects ($2\pi k$-method) was first suggested in \cite{6} and discussed in \cite{3,4,8}. But this method has never been verified for a quantum computer algorithm.

2. A $2\pi k$-method allows one to decrease the total time of computation by taking the maximum possible value of $\Omega$. The minimum time of computation was achieved at the values: $\Omega=2J/\sqrt{15}$ for $\pi/2$-pulses, and $\Omega=2J/\sqrt{3}$ for $\pi$-pulses. The probability to measure an error state at these values of $\Omega$ was of the order $10^{-2}$ (Fig. 13).

3. The probability of error caused by non-resonant effects or by random fluctuation of parameters is connected with the relatively small amount of states. As a rule, one, two, or three error states have the probability of excitation much greater than all other error states taken together. Normally, the upper state has a large probability of excitation.

The next step should include bigger number of qubits, and application of $2\pi k$-method together with error correction approaches.\\ \ \\
{\bf Acknowledgments}\\ \ \\
This work  was supported by the Department of Energy under contract W-7405-ENG-36, and by the National Security Agency.
\newpage

\newpage
\quad\\
Fig. 1:~ Probabilities of the states $|n\rangle$, $|c_n|^2$, ($n=0,\dots,15$) at the end of the Shor's 
algorithm. The parameters are: $\Delta\omega=10$, $J=1.0$ and  $\Omega=0.1$; (a) the probabilities for the ``resonant'' states, with complex amplitudes, $c_1$, $c_3$, $c_5$ and $c_7$; (b) the probabilities for other states (which remain small), at different scale.\\ \ \\
Fig. 2:~ The same as in Fig. 1, for $\Omega=0.112$.\\ \ \\
Fig. 3:~ The same as in Fig. 1, for $\Omega=0.125$.
\\ \ \\
Fig. 4:~ The same as in Fig. 1, for $\Omega=0.1458$.\\ \ \\
Fig. 5:~ The same as in Fig. 1, for $\Omega=0.1666$.\\ \ \\
Fig. 6:~ The same as in Fig. 1, for $\Omega=0.2083$.\\ \ \\
Fig. 7:~ The same as in Fig. 1, for $\Omega=0.2083$.\\ \ \\
Fig. 8:~ Probabilities, $|c_1|^2$, $|c_3|^2$, $|c_5|^2$ and $|c_7|^2$ at the end of the Shor's 
algorithm, as a function of  the Rabi frequency, $\Omega$ . The  parameters are: $\Delta\omega=10$ and $J=1.0$.\\ \ \\
Fig. 9:~ Probabilities, $|c_n|^2$, ($n=0,\dots,15$) at the end of the Shor's 
algorithm, for different values of $\Delta\omega$. The parameters are:
$\Omega=0.1$ and $J=1.0$.\\ \ \\
Fig. 10:~ Dynamics of the probability, $|c_3(t)|^2$, near the end of the Shor's algorithm, for different values of $\Delta\omega$. The parameters are: $J=1$ and  $\Omega=0.1$. Arrows show the end of 14-th, 15-th, and 16-th pulses.\\ \ \\
Fig. 11:~ Probabilities, $|c_n|^2$, ($n=0,\dots,15$) at the end of the Shor's 
algorithm, for random phases: $\varphi_k=\varphi_k^{(0)}+\Delta\varphi_k$;
$\varphi_k^{(0)}$ are the required phases of the pulses ($k=1,...,16$); $\Delta\varphi_k$ are random phases which were chosen in the interval: $\Delta\varphi\in[-\epsilon,\epsilon]$; (1) $\epsilon=0$, (2) $\epsilon=0.1$, (3) $\epsilon=0.5$, (4) $\epsilon=0.8$. The parameters are: $\Delta\omega=10$, $\Omega=0.1$ and $J=1.0$.\\ \ \\
Fig. 12:~ Probabilities, $|c_n|^2$, ($n=0,\dots,15$) at the end of the Shor's 
algorithm, for random duration of resonant pulses, $\tau_n$, ($n=1,...,16$);
$\tau_n=\tau^{(0)}_n+\Delta\tau_n$, where $\tau^{(0)}_n$ is a required duration of the $n$-th pulse, and $\Delta\tau_n$ is a random variable uniformly distributed between $-\epsilon$ and $+\epsilon$. The parameters are: $\Delta\omega=10$, $\Omega=0.1$ and $J=1.0$.\\ \ \\
Fig. 13:~ Probabilities, $|c_n|^2$, $n=0,\dots,15$, at the end of the 
Shor algorithm, for four qubits. The parameters are: $\Delta\omega=10$, 
$J=1.0$, $\Omega=2/\sqrt{3}$ for $\pi$-pulses, and $\Omega=2/\sqrt{15}$ for 
$\pi/2$-pulses.\\ \ \\

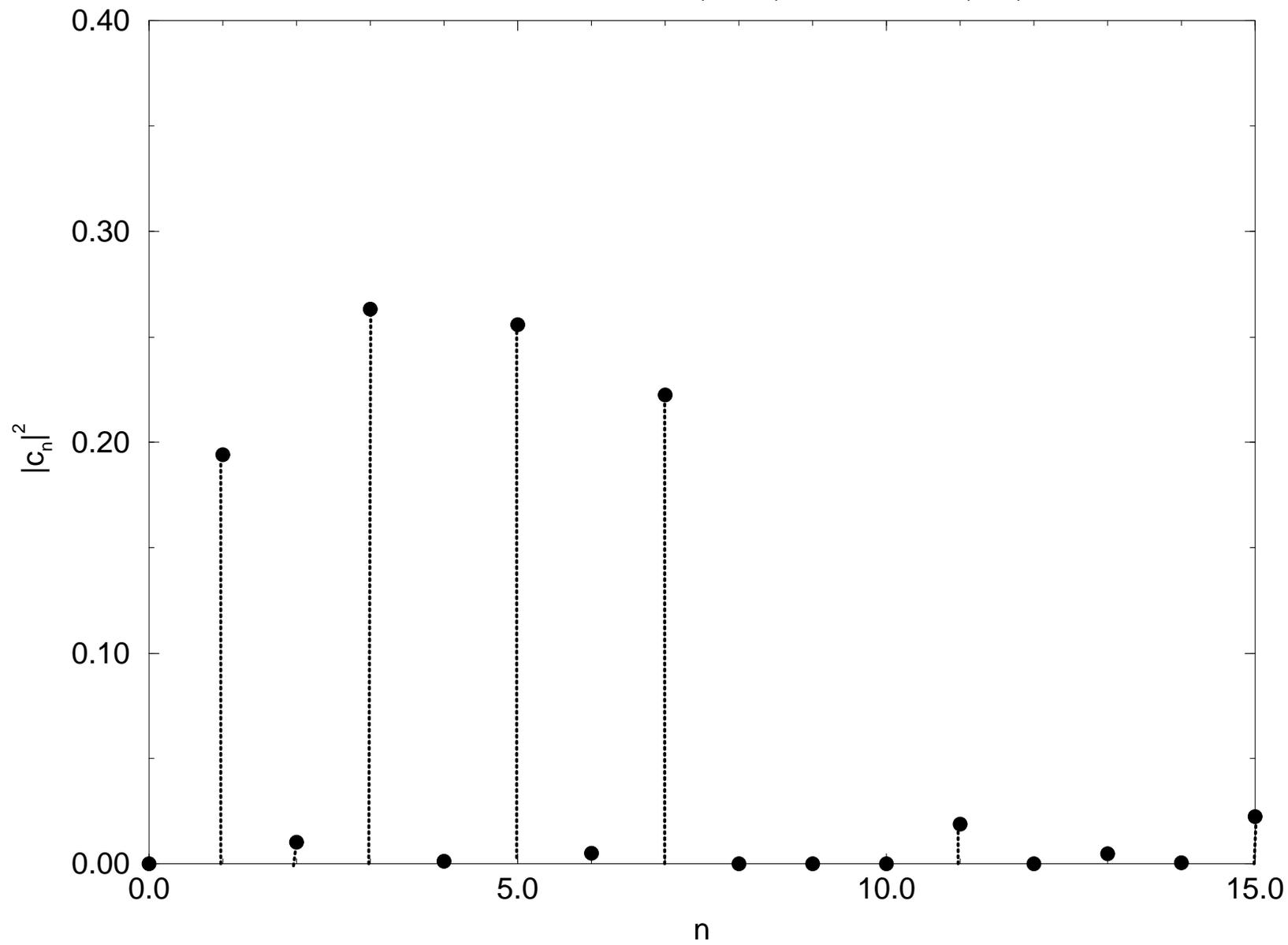
\end{document}